\documentclass[aps,pra,showpacs,superscriptaddress,twocolumn,
amsmath,amssymb,longbibliography]{revtex4-1}
\usepackage{amsfonts}
\usepackage{txfonts}
\usepackage{amsmath}
\usepackage{IEEEtrantools}
\usepackage{float}
\usepackage[colorlinks,breaklinks,linkcolor=blue,anchorcolor=blue,citecolor=blue,urlcolor=blue
]{hyperref}

\usepackage{graphicx}
\usepackage{dcolumn}
\usepackage{bm}
\usepackage{amssymb}
\usepackage{mathrsfs}
\usepackage{graphicx}
\usepackage[sort&compress]{natbib}
\usepackage{subcaption}
\usepackage{caption}
\begin{document}
\title{Non-Hermitian effects on the quantum parameter estimation in pseudo-Hermitian systems}
\author{L. H. Wei}
\affiliation{Center for Quantum Sciences and School of Physics, Northeast Normal University, Changchun 130024, China}
\author{H. J. Xing}
\affiliation{Center for Quantum Sciences and School of Physics, Northeast Normal University, Changchun 130024, China}
\author{L. B. Fu}
\email[]{lbfu@gscaep.ac.cn}
\affiliation{Graduate School of China Academy of Engineering Physics,
No. 10 Xibeiwang East Road, Haidian District, Beijing, 100193, China}
\author{H. D. Liu}
\email[]{liuhd100@nenu.edu.cn}
\affiliation{Center for Quantum Sciences and School of Physics, Northeast Normal University, Changchun 130024, China}
\date{\today}

\begin{abstract}
Quantum Fisher Information (QFI) is a fundamental quantity in quantum parameter estimation theory, characterizing the ultimate precision bound of parameter estimation.
In this work, we investigate  QFI for quantum states in non-Hermitian systems. By employing the projected Hilbert space method and spectral decomposition, we derive an explicit expression for the QFI in terms of the density matrix and parameter generators. This formulation not only recovers the well-known results in the Hermitian case but also captures the non-Hermitian effects induced by the time-dependent norm of the state. To validate our theoretical framework, we analyze a single-qubit pseudo-Hermitian system and apply Naimark dilation theory to embed it into an equivalent Hermitian system. The comparison between the original and dilated systems demonstrates the consistency and applicability of the proposed QFI formula in non-Hermitian settings. In addition, we investigate a $\mathcal{PT}$-symmetric system to further explore the influence of non-Hermiticity on QFI. Our findings offer a new perspective for analyzing and enhancing QFI in non-Hermitian systems, paving the way for promising applications in non-Hermitian quantum metrology and sensing.
\end{abstract}
\maketitle

\allowdisplaybreaks
\section{Introduction}
Quantum parameter estimation theory provides a theoretical framework for acquiring the ultimate precision of estimating parameters \cite{PhysRevLett.102.253601,PhysRevLett.82.4619,RevModPhys.86.121}. As a central metric for quantifying the estimation precision, the quantum Fisher information (QFI) has significantly advanced the development of precision measurement techniques. It has found extensive applications in cutting-edge fields such as quantum sensing \cite{nonmarkovianitylocal,PhysRevA.108.042612,PhysRevA.96.012117,PhysRevA.109.062611,PhysRevLett.112.120405}, quantum imaging \cite{PhysRevLett.117.190802,unknown,PhysRevLett.101.253601,10.3389fphy.2022}, and quantum computation \cite{article20070977,article20211002,unknown202203,dai2023quantumbayesianoptimization}. In classical parameter estimation theory, Fisher Information and the Cram{\'e}r-Rao bound \cite{Cramer,Rao1992} play an important role. Later, Helstrom \cite{Helstrom1969QuantumDA} extended these concepts to the quantum domain, establishing QFI as a fundamental tool to evaluate the accuracy of parameter estimation. While classical Fisher information quantifies the amount of information obtainable from a specific measurement, QFI represents the ultimate bound on information that can be achieved by optimizing all possible quantum measurement strategies.

Although a large number of studies have proposed different calculation methods of the QFI\cite{Braunstein1994,Liu2014,stevenson2018classical}, in complex quantum systems, especially in high-dimensional, non-Hermitian, or mixed-state systems, the calculation of the QFI still faces enormous challenges. While these theoretical assumptions are suitable for closed systems and idealized scenarios, they fall short in describing open quantum systems and non-equilibrium quantum many-body systems\cite{Mori2019Prethermalization,Heyl2021DQPT,Dalla2020QuantumCriticality,Carleo2020BoseHubbard,Perfetto2019Reconstructing,Lesanovsky2021Resetting,Rout2020SpinDynamics}, where many physical characteristics necessitate the adoption of a Non-Hermitian framework \cite{Li_2019,Wu_2019,Naghiloo_2019,Lin_2022,Xiao_2020}. Therefore, it becomes imperative to refine the theoretical approach and investigate QFI within the context of non-Hermitian quantum systems.

Non-Hermitian systems, including pseudo-Hermitian and $\mathcal{PT}$-symmetric systems, are gradually attracting widespread attention due to their unique advantages in open quantum systems and topological quantum physics \cite{FHMFaisal1983,Mostafazadeh2002,Mosta2004,Bagchi_2002,Mostafazadeh_2002,Mostafazadeh2004}. In recent years, research on QFI in non-Hermitian, particularly pseudo-Hermitian systems, has gradually progressed, providing a theoretical foundation for high-precision quantum sensing and parameter estimation \cite{PhysRevA.110.012413,PhysRevB.110.045202}. The QFI of pseudo-Hermitian systems can not only reveal the sensitivity of the system to external perturbations but can also be extended to equivalent Hermitian systems through the Naimark dilation theory \cite{PhysRevLett.131.160801}, enabling the calculation and optimization of QFI. This method retains the nonorthogonal state characteristics of non-Hermitian systems, captures coherent effects in complex evolutions, and enhances the precision of parameter estimation. However, other aspects of non-Hermitian features, such as the non-normalized norm of the state, have not yet been thoroughly investigated.

This study aims to explore the non-Hermitian effects on the calculation and optimization of QFI in non-Hermitian and pseudo-Hermitian systems. By deriving an explicit expression for the QFI in terms of the density matrix and parameter generators in non-Hermitian systems. We investigate the role of non-Hermiticity in parameter estimation within pseudo-Hermitian systems. A single-qubit pseudo-Hermitian system and a $\mathcal{PT}$-symmetric system to demonstrate the feasibility of the proposed formulas and investigate the non-Hermitian effects on quantum parameter estimation.  The research indicates that pseudo-Hermitian and $\mathcal{PT}$-symmetric systems have broad prospects for improving measurement precision in quantum metrology, bringing new opportunities to non-Hermitian physics and quantum sensing fields.

This paper is organized as follows. In Sec. \ref{sec2}, we first present the expression of QFI based on density matrices and parameter generators in non-Hermitian systems. In Sec. \ref{sec3}, we apply this formulation to analyze the QFI in pseudo-Hermitian systems and demonstrate that the Naimark dilation to an equivalent Hermitian system more effectively preserves parameter information.  In Sec. \ref{sec4}, we calculate the QFI using parameter generators in a $\mathcal{PT}$-symmetric system.  By comparing the unbroken $\mathcal{PT}$-symmetric region and the broken $\mathcal{PT}$-symmetry regions, the analysis identifies the optimal initial states most suitable for parameter estimation. Finally, Sec. \ref{sec5} provides a summary and concluding remarks for the paper.

\section{THE FISHER INFORMATION IN THE PROJECTED HILBERT SPACE OF NON-HERMITIAN SYSTEMS}\label{sec2}
It is well known that the inverse of Fisher information provides a lower bound on the accuracy limit of parameter estimation \cite{Li2021QFI,Cieslinski2024ManyBodyQFI,QuantumMetrology2016}. Correspondingly, in quantum metrology, QFI represents the maximum Fisher information over all possible measurement schemes \cite{Braunstein1994}. The QFI with respect to the parameter $\theta$ in Hermitian systems is defined as
\begin{equation}\label{Eq.1}
    F\equiv\langle\hat{L}^{2}\rangle,
\end{equation}
where $\hat{L}$ is the symmetric logarithmic derivative operator\cite{jiang2014quantum,zhang2023direct,chou2020symmetric}, which satisfies
\begin{equation}
    \partial_{\theta}\hat{\rho}_{\theta}=(\hat{L}\hat{\rho}_{\theta}+\hat{\rho}_{\theta}\hat{L})/2
\end{equation}
 with $\rho_\theta$ being  the density matrix of a mixed state with parameter $\theta$. In the case of pure states $|\psi_\theta\rangle$ , the density matrix simplifies to $\hat{\rho}_{\theta}=|\psi_{\theta}\rangle\langle\psi_{\theta}|$. Thus, the QFI for pure states in a Hermitian system is given by \cite{Liu2014}
	\begin{equation}\label{Eq.2}
		\begin{aligned}
			F&=\mathrm{tr}(\hat{\rho}_{\theta}\hat{L}^{2})=\sum_{k}\langle\psi_{\theta}|\hat{L}|k_{\theta}\rangle\langle k_{\theta}|\hat{L}|\psi_{\theta}\rangle\\
   &=4\left(\langle\partial_{\theta}\psi_{\theta}|\partial_{\theta}\psi_{\theta}\rangle+\langle\partial_{\theta}\psi_{\theta}|\psi_{\theta}\rangle^{2}\right).\\
		\end{aligned}
	\end{equation}
Assuming that the parameter $\theta$ is related to a unitary evolution acting on an initial state, its unitary operator  $\hat{U}(\theta)$ takes the form
\begin{equation}\label{Eq.3}
    \hat{U}(\theta)=\exp(i\theta\mathcal{H}),
\end{equation}
where $\mathcal{H}$ is the generator of parameter $\theta$. That initial state, which is independent of the parameter, is represented as
\begin{equation}\label{Eq.4}
    \hat{\rho}_{0}=|\psi_{0}\rangle\langle\psi_{0}|.
\end{equation}
Following the unitary evolution, the state becomes
\begin{equation}\label{Eq.5}
    \hat{\rho_{\theta}}=\hat{U}_{\theta}\hat{\rho_{0}}\hat{U}^{\dagger}_{\theta}
\end{equation}
Substitute  $\mathcal{H}=i(\partial_{\theta}\hat{U}^{\dagger}_{\theta})\hat{U}_{\theta}$ into Eq.(\ref{Eq.2}), $F_\theta$ can be rewritten as
\begin{equation}\label{Eq.6}
    F_\theta=4\langle\psi_{0}|\Delta\mathcal{H}^{2}|\psi_{0}\rangle,
\end{equation}
where $\Delta\mathcal{H}$ is the variance of Hamiltonian on the state $|\psi_0\rangle$.

In the non-Hermitian case, we begin by applying the Schwarz inequality \cite{Schwarz1865} to study QFI in non-Hermitian systems, similar to what has been done in the Hermitian case. The Schwartz inequality
\begin{equation}\label{Eq.7}
\langle\varphi|\varphi\rangle\langle\psi|\psi\rangle\geq\langle\varphi|\psi\rangle\langle\psi|\varphi\rangle
\end{equation}
for any two states $|\varphi\rangle$ and $|\psi\rangle$ holds for any two states regardless whether the states are normalized. However,  if we set $|\psi\rangle=\hat{A}|\phi\rangle$ and $|\varphi\rangle=\hat{B}|\phi\rangle$, unlike in the Hermitian case, the relationship between non-Hermitian operators $\hat{A}$ and $\hat{B}$ becomes
\begin{equation}\label{Eq.8}
\langle\hat{A}^{\dagger}\hat{A}\rangle\langle\hat{B}^{\dagger}\hat{B}\rangle\geq\langle\hat{B}^{\dagger}\hat{A}\rangle\langle\hat{A}^{\dagger}\hat{B}\rangle.
\end{equation}
Then, the Robertson-Schr{\"o}dinger inequality for non-Hermitian operators takes the form
\begin{multline}\label{Eq.9}
\langle\Delta\hat{A}^{\dagger}\Delta\hat{A}\rangle\langle\Delta\hat{B}^{\dagger}\Delta\hat{B}\rangle \geq\\ \left(\frac{\langle\hat{A}^{\dagger}\hat{B}+\hat{B}^{\dagger}\hat{A}\rangle}{2}-\frac{\langle\hat{A}\rangle^{*}\langle\hat{B}\rangle+\langle\hat{A}\rangle\langle\hat{B}\rangle^{*}}{2}\right)^{2}\\
		+\left|\frac{\langle\hat{A}^{\dagger}\hat{B}-\hat{B}^{\dagger}\hat{A}\rangle}{2}-\frac{\langle\hat{A}\rangle^{*}\langle\hat{B}\rangle-\langle\hat{A}\rangle\langle\hat{B}\rangle^{*}}{2}\right|^{2}.
\end{multline}
Therefore, when measuring a Hermitian operator in a non-Hermitian system, one can utilize the above formula to obtain the uncertainty relation between the Hermitian operator $\hat{A}$ and the non-Hermitian operator $\hat{L}$ as
\begin{multline}\label{Eq.10}
		\langle\Delta\hat{A}\rangle^{2}(\langle\hat{L}^{\dagger}\hat{L}\rangle-{\langle\hat{L}\rangle}^{*}\langle\hat{L}\rangle)^{2}\\
 \geq\left(\frac{\langle\hat{A}\hat{L}+\hat{L}^{\dagger}\hat{A}\rangle}{2}-\frac{\langle\hat{A}\rangle\langle\hat{L}\rangle+\langle\hat{A}\rangle{\langle\hat{L}\rangle}^{*}}{2}\right)^{2}\\
		+\left|\frac{\langle\hat{A}\hat{L}-\hat{L}^{\dagger}\hat{A}\rangle}{2}-\frac{\langle\hat{A}\rangle\langle\hat{L}\rangle-\langle\hat{A}\rangle{\langle\hat{L}\rangle}^{*}}{2}\right|^{2}.
\end{multline}
To tighten the inequality, we introduce a non-Hermitian operator $\hat{L}$ with an expectation value of $0$ and satisfies $\hat{A}\hat{L}=\hat{L}^{\dagger}\hat{A}$. This choice enhances the precision of the parameter being measured. By substituting the operator $\hat{L}$ into the error propagation formula $\langle\Delta\hat{A}\rangle^{2}=\delta\theta^{2}{\partial_{\theta}\langle\hat{A}\rangle|_{\bar{\theta}}}^{2}$, we derive inequalities of the form
\begin{equation}\label{Eq.11}
\delta\theta^{2}\geq\dfrac{1}{\langle\hat{L}^{\dagger}\hat{L}\rangle}\cdot\dfrac{\langle\hat{A}\hat{L}+\hat{L}^{\dagger}\hat{A}\rangle^{2}}{4{\partial_{\theta}\langle\hat{A}\rangle|_{\bar{\theta}}}^{2}}.
\end{equation}
Using the properties of the density matrix, we obtain
\begin{subequations}
	\begin{equation}\label{Eq.12a}
		\mathrm{tr}\left(\hat{A}\hat{L}\hat{\rho}+\hat{L}^{\dagger}\hat{A}\hat{\rho}\right)=\mathrm{tr}\left(\hat{A}(\hat{L}\hat{\rho}+\hat{\rho}\hat{L}^{\dagger})\right),
	\end{equation}
	\begin{equation}\label{Eq.12b}
		\partial_{\theta}\langle\hat{A}\rangle=\partial_{\theta}\mathrm{tr}(\hat{\rho}\hat{A})=\mathrm{tr}(\hat{A}\partial_{\theta}\hat{\rho}).
	\end{equation}
\end{subequations}
Let $\partial_{\theta}\hat{\rho}=\frac{1}{2}(\hat{L}\hat{\rho}+\hat{\rho}\hat{L}^{\dagger}),$ then Eq.(\ref{Eq.11}) becomes
\begin{equation}\label{Eq.13}
	\delta\theta^{2}\geq\dfrac{1}{\langle\hat{L}^{\dagger}\hat{L}\rangle}.
\end{equation}
Consequently, the QFI can be expressed as
\begin{equation}
	F_\theta=\langle\hat{L}^{\dagger}\hat{L}\rangle. \label{Eq.14}
\end{equation}

The distinction between Hermitian and non-Hermitian systems lies in the fact that the norm of a general state $|\Psi\rangle$ in non-Hermitian systems is normally time-dependent. As a result, the trace of the density matrix $\rho=|\Psi\rangle\langle\Psi|$ is not conserved, which affects the expectation value $\langle\hat{A}\rangle=\langle\Psi|\hat{A}|\Psi\rangle$. To analyze this non-Hermitian behavior,  we employ the method of projective Hilbert space for non-Hermitian systems \cite{Fu_2024,PhysRevA.87.013629}. Specifically, $|\Psi\rangle$ can be decomposed as
\begin{equation}
    |\Psi\rangle=e^{\alpha+i\beta}|\psi\rangle,
\end{equation}
where $|\psi\rangle$ is normalized, i.e., $\langle\psi|\psi\rangle=1$. Here, $\alpha$ and $\beta$ are two real numbers whose time evolution is governed by
\begin{equation}
    \dot\alpha=-\frac i 2\langle\psi|\hat H-\hat H^\dagger|\psi\rangle, ~~~\dot\beta=-\frac 1 2\langle\psi|\hat H+\hat H^\dagger|\psi\rangle+i\langle\psi|\dot\psi\rangle.
\end{equation}
These equations describe the influences of  the non-Hermitian components of the Hamiltonian and the phase factors, respectively.
For the density matrix $\hat{\rho}_{\theta}$, its trace is influenced by $\alpha$, i.e. $\hat{\rho}_{\theta}=|\Psi_{\theta}\rangle\langle\Psi_{\theta}|=e^{2\alpha}|\psi_{\theta}\rangle\langle\psi_{\theta}|$, its related matrix elements in terms of the projective state $|\psi_\theta\rangle$ are given by
	\begin{subequations}
	   \begin{equation}\label{Eq.15a}
	       \langle k|\partial_{\theta}\hat{\rho}_{\theta}|\psi_{\theta}\rangle=\frac{e^{2\alpha}}{2}\langle k_{\theta}|\hat{L}|\psi_{\theta}\rangle,
	   \end{equation}
	\begin{equation}\label{Eq.15b}
\langle\psi_{\theta}|\partial_{\theta}\hat{\rho}_{\theta}|k_{\theta}\rangle=\frac{e^{2\alpha}}{2}\langle\psi_{\theta}|\hat{L}^{\dagger}|k_{\theta}\rangle.
	\end{equation}
	\end{subequations}
Substitute these into Eq. (\ref{Eq.14}), we obtain the QFI for pure states in non-Hermitian systems as
\begin{equation}\label{Eq.16}
\begin{aligned}    F&=\sum_{k}e^{2\alpha}\langle\psi_{\theta}|\hat{L}^{\dagger}|k_{\theta}\rangle\langle k_{\theta}|\hat{L}|\psi_{\theta}\rangle\\
    &=\sum_{k}\frac{4}{e^{2\alpha}}\langle\psi_{\theta}|\partial_{\theta}\hat{\rho}_{\theta}|k_{\theta}\rangle\langle k_{\theta}|\partial_{\theta}\hat{\rho}_{\theta}|\psi_{\theta}\rangle\\
    &=16e^{2\alpha}(\partial_{\theta}\alpha)^{2}+4e^{2\alpha}[\langle\partial_{\theta}\psi_{\theta}|\partial_{\theta}\psi_{\theta}\rangle-|\langle\partial_{\theta}\psi_{\theta}|\psi_{\theta}\rangle|^{2}].
\end{aligned}
\end{equation}
It is evident that this QFI for non-Hermitian systems comprises two contributions: one from the projective state $|\psi_\theta\rangle$ similar to the Hermitian system, and the other from the non-Hermitian effects brought by  $\alpha$.

When describing the QFI using the parameter generator, the non-Hermiticity parameter $\alpha$  causes the final state $|\Psi_{\theta}\rangle$ to lose normalization during the evolution. Therefore the evolution operator is no longer unitary due to $\mathcal{H}\neq\mathcal{H}^{\dagger}$. However, the project state $|\psi_{\theta}\rangle$ remains normalized throughout the evolution process  \cite{PhysRevA.108.022215}. Assuming the initial state is normalized, i.e.,  $|\Psi_\theta\rangle=e^{\alpha}|\psi_\theta\rangle$,  we can express Eq.(\ref{Eq.16}) as
\begin{equation}\label{Eq.17}
    F_{\theta}=16e^{2\alpha}(\partial_{\theta}\alpha)^{2}+4(\langle\hat{\mathcal{H}^{\dagger}}\hat{\mathcal{H}}\rangle_{\theta}-\langle\hat{\mathcal{H}^{\dagger}}\rangle_{\theta}\langle\hat{\mathcal{H}}\rangle_{\theta}),
\end{equation}
where $\langle\hat{M}\rangle_{\theta}$ denotes the expectation value under a normalized final state $|\psi_{\theta}\rangle$ (i.e., $\langle\hat{\mathcal{H}}\rangle_{\theta}=\langle\psi_{\theta}|\hat{\mathcal{H}}|\psi_{\theta}\rangle$). Since $|\Psi_{\theta}\rangle$ is not normalized , we define the inner product factor as \begin{equation}P_{\theta}=\langle\Psi_{\theta}|\Psi_{\theta}\rangle=\langle\psi_{0}|\hat{U}_{\theta}^{\dagger}\hat{U}_{\theta}|\psi_{0}\rangle=e^{2\alpha}.
\end{equation}
It is evident that the QFI varies with the parameter $\theta$ as well as the non-Hermicity parameter $\alpha$. If the Hamiltonian is Hermitian, this result returns to the Hermitian case, where $F_{\theta}=4\langle\psi_{0}|\Delta\hat{\mathcal{H}}^{2}|\psi_{0}\rangle$.

The results we derived for the pure states in the non-Hermitian system can be easily extended to the case of mixed states. Similar to the decomposition of the state
\begin{equation}
    |\psi_{\theta}\rangle=e^{\alpha+i\beta}|\psi_\theta\rangle=e^{\alpha+i\beta}\sum_{i}^{N}C_{i}|i_{\theta}\rangle,
\end{equation}
the density matrix of the mixed state can be expressed as
\begin{equation}
\hat{\rho}_{\theta}=e^{2\alpha}\sum_{i}^{M}p_{i}|i_{\theta}\rangle\langle i_{\theta}|.
\end{equation}
Here, $p_i$ represents the probability associated with the state $|i_\theta\rangle$, and  $M$ is the rank of the density matrix, which may be less than the dimension $N(\geq M)$ of the Hilbert space spanned by  $|\psi_{\theta}\rangle$. This treatment in the projective Hilbert space offers significant advantages. Although the inner product  $P_\theta$ and the trace of $\hat\rho_\theta$ are not normalized during an evolution process, the total probability $\sum_i|C_i|^2$ and $\sum_ip_i$ are conserved. This ensures consistency in the probabilistic interpretation of the system, even in the non-Hermitian framework.

The matrix elements for $\partial_{\theta}\hat{\rho}$ are given by
	\begin{subequations}
 \begin{equation}\label{Eq.18a}
     \langle l_{\theta}|\partial_{\theta}\hat{\rho}_{\theta}|k_{\theta}\rangle=e^{2\alpha}p_{k}\langle l_{\theta}|\hat{L}|k_{\theta}\rangle,
 \end{equation}
 \begin{equation}\label{Eq.18b}
     \langle k_{\theta}|\partial_{\theta}\hat{\rho}_{\theta}|l_{\theta}\rangle=e^{2\alpha}p_{k}\langle k_{\theta}|\hat{L}^{\dagger}|l_{\theta}\rangle.
 \end{equation}	
	\end{subequations}
Substituting these into Eq.(\ref{Eq.14}), we obtain
\begin{equation}\label{Eq.19}
F = \langle \hat{L}^{\dagger}\hat{L} \rangle = e^{-2\alpha}\sum_{k=1}^{M}\sum_{l=1}^{N}\frac{2}{p_{k}+p_{l}}|(\partial_{\theta}\hat{\rho}_{\theta})_{kl}|^{2}.
\end{equation}
The summation in the above equation includes probabilities $p_{k}$ and $p_{l}$, where both $l=k$ and $l\neq k$ are possible cases. The asymmetric treatment yields
\begin{equation}\label{Eq.20}
\begin{aligned}   F&=e^{-2\alpha}\left[\sum_{k=1}^{M}\sum_{l=1}^{N}\frac{4p_{k}}{(p_{k}+p_{l})^{2}}|(\partial_{\theta}\hat{\rho}_{\theta})_{kl}|^{2}\right.\\
   &\left.+\sum_{k=1}^{M}\sum_{l=1}^{N}\frac{4}{p_{k}}|(\partial_{\theta}\hat{\rho}_{\theta})_{kl}|^{2} \right] .
\end{aligned}
\end{equation}
Using the definition of mixed states in the projective Hilbert space, we derive the partial derivative of $\hat{\rho}_{\theta}$ with respect to the parameter. Considering $\sum_{k=1}^{M}\sum_{l=M+1}^{N}|\langle\partial_{\theta}k_{\theta}|l_{\theta}\rangle|^{2}=\sum_{k=1}^{M}\langle\partial_{\theta}k_{\theta}|\partial_{\theta}k_{\theta}\rangle-\sum_{k=1}^{M}\sum_{l=1}^{M}|\langle\partial_{\theta}k_{\theta}|l_{\theta}\rangle|^{2}$, we obtain the QFI for mixed states as
\begin{equation}\label{Eq.21}	
\begin{aligned}
F=&\sum_{k=1}^{M}\frac{4e^{2\alpha}(\partial_{\theta}p_{k})^{2}}{p_{k}}+\sum_{k=1}^{M}4p_{k}e^{2\alpha}(\langle\partial_{\theta}k_{\theta}|\partial_{\theta}k_{\theta}\rangle-|\langle\partial_{\theta}k_{\theta}|k_{\theta}\rangle|^{2})\\&-\sum_{k\neq l}^{M}e^{2\alpha}\frac{8p_{k}p_{l}}{p_{k}+p_{l}}|\langle\partial_{\theta}k_{\theta}|l_{\theta}\rangle|^{2}\\
	&+\sum_{k=1}^{M}[16p_{k}e^{2\alpha}(\partial_{\theta}\alpha)^{2}+8e^{2\alpha}\partial_{\theta}p_{k}\partial_{\theta}\alpha].
\end{aligned}
\end{equation}
The QFI formulas above are all calculated under non-Hermitian conditions. However, when the system is Hermitian, the setting $\alpha=0$ reduces the expressions to the Hermitian case.

 Since $p_{k}$ is conserved under unitary evolution, the QFI for the mixed state $\hat{\rho_{\theta}}$ becomes
\begin{equation}\label{Eq.22}
\begin{aligned}
   F_{\theta}&=\sum_{k=1}^{M}4p_{k}e^{2\alpha}(\langle\partial_{\theta}k_{\theta}|\partial_{\theta}k_{\theta}\rangle-|\langle\partial_{\theta}k_{\theta}|k_{\theta}\rangle|^{2})\\
   &-\sum_{k\neq l}^{M}e^{2\alpha}\frac{8p_{k}p_{l}}{p_{k}+p_{l}}|\langle\partial_{\theta}k_{\theta}|l_{\theta}\rangle|^{2}
	+\sum_{k=1}^{M}16p_{k}e^{2\alpha}(\partial_{\theta}\alpha)^{2}.
\end{aligned}
\end{equation}
Using the parameter operator, the density matrix $\hat{\rho}_{\theta}$ is expressed as
\begin{equation}\label{Eq.23}
\hat{\rho}_{\theta}=\hat{U}_{\theta}\hat{\rho}_{0}\hat{U}_{\theta}^{\dagger}=\sum_{i}^{d}p_{i}\hat{U}_{\theta}|i_{0}\rangle\langle i_{0}|\hat{U}_{\theta}^{\dagger},
\end{equation}
where the $p_{i}$ is the $i$th eigenvalue of the density matrix $\hat\rho_0$, $|i_{0}\rangle$ is the corresponding eigenstate, and $d(d\leq N)$ is the dimension of $\hat{\rho}_{0}$. It is easy to notice that $p_{i}$ and $\hat{U}_{\theta}|i_{0}\rangle$ are the corresponding eigenvalue and eigenstate of $\hat{\rho}_{\theta}$ \cite{Liu2018QuantumFI}. The QFI of $\hat{\mathcal{H}}$ is given by
\begin{equation}\label{Eq.24}
    \begin{aligned}
       F_{\theta}&= \sum_{k=1}^{M}4p_{k}e^{2\alpha}(\langle k_{\theta}|\hat{\mathcal{H}}^{\dagger}\hat{\mathcal{H}}|k_{\theta}\rangle-\langle k_{\theta}|\hat{\mathcal{H}}^{\dagger}|k_{\theta}\rangle\langle k_{\theta}|\hat{\mathcal{H}}|k_{\theta}\rangle)\\
   &-\sum_{k\neq l}^{M}\frac{8e^{2\alpha}p_{k}p_{l}}{p_{k}+p_{l}}|\langle k_{\theta}|\hat{\mathcal{H}}^{\dagger}|l_{\theta}\rangle|^{2}
	+\sum_{k=1}^{M}16p_{k}e^{2\alpha}(\partial_{\theta}\alpha)^{2}.
    \end{aligned}
\end{equation}
In summary, the effects of non-Hermiticity on the QFI for both pure and mixed states in non-Hermitian systems are caused by the time-dependent norm $e^{2\alpha}$ of the states. Notably, these results seamlessly reduce to the Hermitian case when  $\alpha=0$, demonstrating the consistency of the formalism with established results in Hermitian quantum mechanics.


\section{SINGLE-QUBIT PSEUDO-HERMITIAN SYSTEM}\label{sec3}
\setlength{\parskip}{0.1\baselineskip}
To illustrate our results, we consider a simple two-level system characterized by a pseudo-Hermitian Hamiltonian  \cite{Mostafazadeh2002,PhysRevLett.124.020501}, given by
	\begin{equation}\label{Eq.25}
		\hat{H}_{s}=\varepsilon
		\left(
		\begin{array}{cc}
			0 & \delta^{-1} \\
			\delta & 0
		\end{array}
		\right),
	\end{equation}
which possesses real eigenvalues $\pm \varepsilon$. This pseudo-Hermitian quantum system satisfies the Schrödinger equation $i\partial_{t}|\psi(t)\rangle_{s}=\hat{H}_{s}|\psi(t)\rangle_{s}$.

To gain a clearer physical understanding of information retrieval and to clarify the relation between the QFI in a non-Hermitian system and  the QFI in a related Hermitian system, we consider embedding this pseudo-Hermitian system into a larger Hermitian system.
By adding an ancilla (a measuring apparatus) and extending the Hilbert space, any nonunitary dynamics can be understood as unitary dynamics of the entire system followed by quantum measurement acting on the ancilla. We establish the large Hermitian system in two different ways.

First, using the Naimark dilation theory, the non-unitary dynamics of the non-Hermitian system are mapped onto unitary dynamics within the extended Hilbert space. This Naimark dilation process involves a mapping $\mathcal{M}:\hat{\mathcal{H}}\rightarrow \hat{H}_{s}$, which extends the pseudo-Hermitian dynamics $|\psi(t)\rangle_{s}$ of the system by introducing  the ancilla qubit  state $|0\rangle_{a}$. The time evolution of the entire system governed by the dilated Hamiltonian is given by
\begin{equation}    |\Psi(t)\rangle=|0\rangle_{a}\otimes|\psi(t)\rangle_{s}+|1\rangle_{a}\otimes|\chi(t)\rangle_{s}.
\end{equation}
For a pseudo-Hermitian Hamiltonian, the relation \begin{equation}
    (\eta^{2}+\mathbb{I})\hat{H}_{s}=\hat{H}_{s}^{\dagger}(\eta^{2}+\mathbb{I})
\end{equation}
holds, where $\eta$ is a positive Hermitian operator, ensuring the Hermiticity of the dilated Hamiltonian $\hat{\mathcal{H}}$.  It is also noted that there exists the relation $|\chi(t)\rangle_{s}=\eta|\psi(t)\rangle_{s}$,

To generalize, consider the detection of a weak field acting on the sensor system, which can be described by $V_{s}=\lambda\hat{\sigma}^{(s)}_{x}$, where $\lambda$ is the parameter to be estimated. The complete two-qubit dilated system is governed by the following total Hamiltonian \cite{PhysRevLett.101.230404}
\begin{equation}\label{Eq.26}
    \hat{H}_{tot}=\hat{\mathcal{H}}+\hat{I}^{(a)}\otimes V_{s},
\end{equation}
where the unperturbed Hamiltonian is given by
\begin{equation}
    \hat{\mathcal{H}}=b\hat{I}^{(a)}\otimes\hat{\sigma}^{(s)}_{x}-c\hat{\sigma}^{(a)}_{y}\otimes\hat{\sigma}^{(s)}_{y}
    \end{equation}
which satisfies the mapping relation  $\mathcal{M}(\hat{\mathcal{H}})=\hat{H}_{s}(\delta)
$. The coefficients are defined as $b\equiv4\omega\varepsilon(1+\varepsilon)/(1+2\varepsilon)$ and $c\equiv2\omega\sqrt{\varepsilon(1+\varepsilon)}/(1+2\varepsilon)$, where $\varepsilon$ and $\omega$ characterize the qubit system. Upon introducing the perturbation field $V_s$, the total Hamiltonian system satisfies the mapping relation  $\mathcal{M}(\hat{H}_{tot})=\hat{H}_{s}(\delta_{\lambda})$. The Hermitian Hamiltonian of the dilated two-qubit system is then expressed as
\begin{equation}\label{Eq.27}
		\hat{H}_{tot}=b\hat{I}^{(a)}\otimes\hat{\sigma}_{x}^{(s)}-c\hat{\sigma}_{y}^{(a)}\otimes\hat{\sigma}_{y}^{(s)}+\lambda\hat{I}^{(a)}\otimes\hat{\sigma}_{x}^{(s)}.
	\end{equation}
Consequently, the reduced pseudo-Hermitian Hamiltonian of the two-qubit system, incorporating the perturbation field and corresponding to Eq. (\ref{Eq.25}), takes the form	
 \begin{equation}\label{Eq.28}
		\hat{H}_{s}=\varepsilon_{\lambda}
		\left(
		\begin{array}{cc}
			0 & \delta_{\lambda}^{-1} \\
			\delta_{\lambda} & 0
		\end{array}
		\right),
	\end{equation}
with eigenvalues $\pm\varepsilon_{\lambda}$, where $\varepsilon_{\lambda}=\sqrt{(b+\lambda)^{2}+c^{2}}$ and $\delta_{\lambda}=(\lambda+2\varepsilon\omega)/\varepsilon_{\lambda}$.

For the pseudo-Hermitian Hamiltonian before dilation, the non-Hermitian Fisher information can be calculated as defined  in Eq. (\ref{Eq.20}). By utilizing Naimark dilation theory, we bypassed the analysis of the complex non-unitary parameter encoding process by projecting the dynamics onto an enlarged system, effectively projecting it onto an equivalent enlarged Hermitian system. It is possible to determine the Fisher information of the dilated Hermitian system. Next, we explore the relationship between the non-Hermitian quantum Fisher information (QFI) for the energy eigenstates and its corresponding Hermitian counterpart.

\begin{figure}[t]
	\centering
	\includegraphics[width=0.45\textwidth]{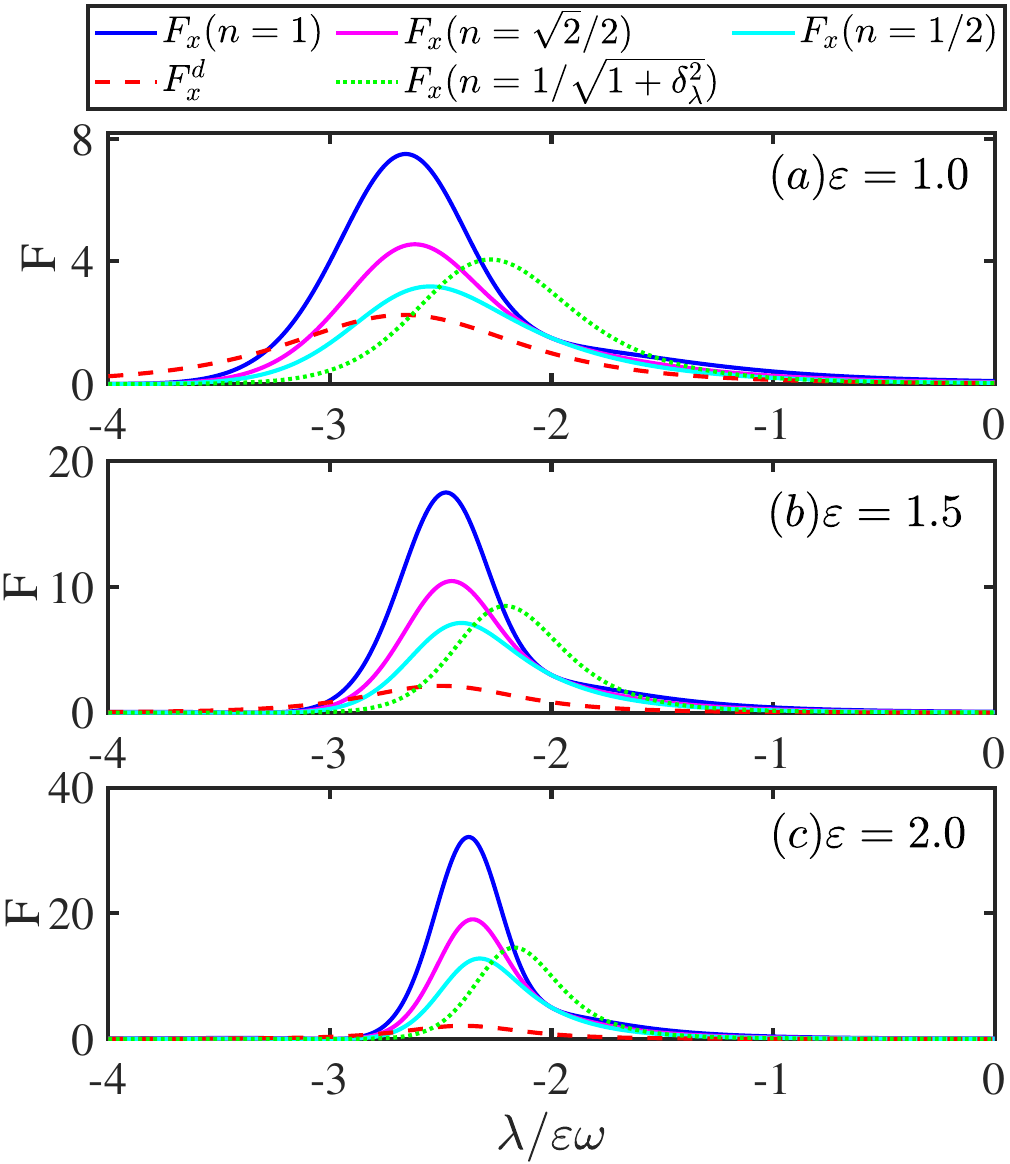}
	\caption{QFI $F_x(n)$ of the state $|R_n\rangle$ as a function of the estimated parameter $x$ for different values of $\varepsilon$: (a) $\varepsilon=1.0$, (b) $\varepsilon=1.5$, (c) $\varepsilon=2.0$. The blue, purple, and cyan solid lines correspond to different choices of  the parameter $n= 1/2, \sqrt{2}/2$ and $1$, respectively. The red dotted line indicates the QFI $F_x^{d}$ obtained after the Naimark dilation, while the green dash-dot line represents the QFI associated with the projected state $|\psi_1\rangle$. The remaining parameter is chosen as $\omega=1$.}
 \label{fig.1}
\end{figure}

For the pseudo-Hermitian Hamiltonian of the two-qubit system, the eigenstate corresponding to the eigenvalue $\varepsilon_{\lambda}$ is given by

     \begin{equation}\label{Eq.29a}
	|R\rangle=n
	\left(
	\begin{array}{c}
		1\\
		\delta_{\lambda}
	\end{array}
\right),
\end{equation}
where $n$ is an arbitrary coefficient, chosen for normalization per the left eigenstate
	\begin{equation}\label{Eq.29b}
		|L\rangle=\frac{1}{2n}
		\left(
		\begin{array}{c}
			1\\
			\delta_{\lambda}^{-1}
			\end{array}
			\right).
		\end{equation}
By considering the right state in the projective Hilbert space as  \begin{equation}
    |R_n\rangle\equiv e^{\alpha_n}|\psi_{1}\rangle
\end{equation}
with the non-Hermticity coefficient $\alpha_n=\ln \left(n\sqrt{1+\delta_\lambda^2}\right)$, the projective state $|\psi_1\rangle =\frac{1}{\sqrt{1+\delta^2_\lambda}}(1\quad \delta_\lambda)^T$. For the dimensionless parameter $x=\lambda/\varepsilon\omega$ to be estimated, we can then derive the quantum Fisher information (QFI) from Eq. (\ref{Eq.16}) as
  \begin{equation}\label{Eq.30}
		F_{x}(n)=\dfrac{4n^{2}g^{2}(4\delta_{\lambda}^{2}+1)}{\delta_{\lambda}^{2}+1}+32ng\delta_{\lambda}(\partial_{\lambda}n)+16(1+{\delta_{\lambda}}^{2})(\partial_{\lambda}n)^2,
	\end{equation}
where $g$ is the partial derivative of $\delta_{\lambda}$ with respect to the parameter to be estimated, i.e., \begin{equation}
g=\partial_{x}\delta_{\lambda}=\dfrac{\varepsilon\omega}{\varepsilon_{\lambda}}-\dfrac{\varepsilon\omega\delta_{\lambda}(b+\lambda)}{\varepsilon_{\lambda}^{2}}.
\end{equation}
It is worth noticing that the value of the QFI depends on the value of   \(n\). For the states $|R\rangle_n$ with different $n$, the resulting QFI values vary (as shown in Fig. \ref{fig.1}). When $n=1/\sqrt{1+\delta_\lambda^2}$, the right state $|R\rangle_n$ correspond to a normalized state $|\psi_1\rangle=\frac{1}{\sqrt2}(|R\rangle_n|0\rangle)$, and now the QFI is
\begin{equation}
    F_x=\frac{4(\partial_x\delta_\lambda)^2}{(1+\delta_\lambda^2)^2}.
\end{equation}
Therefore, to achieve the optimal measurement, it appears crucial to select the appropriate state. In the following, we will compare this result with the corresponding Hermitian system.

 For the Hermitian Hamiltonian expanded according to Naimark dilation theory, when the eigenvalue is $\varepsilon_{\lambda}$, the corresponding eigenstate is
 \begin{equation}\label{Eq.31}
	|\psi_{2}\rangle=\frac{\sqrt{2}}{2}
	\left(
	\begin{array}{c}
		0\\
		1\\
		\frac{c}{\varepsilon_{\lambda}}\\
		\frac{b+\lambda}{\varepsilon_{\lambda}}
	\end{array}
	\right).
\end{equation}
By using Eq. (\ref{Eq.2}), the QFI of eigenstate $|\psi_2\rangle$ is given by
\begin{equation}\label{Eq.32}
	F_{x}^{d}=\frac{2c^{2}}{\varepsilon_{\lambda}^{4}}.
\end{equation}

After deriving the QFI for the non-Hermitian system and the corresponding expressions for the Hermitian systems in Eq.(\ref{Eq.30}) and (\ref{Eq.32}), we can proceed to plot the relationship between the parameter to be estimated and the QFI for the cases where $\varepsilon=1,1.5,2$ and $\omega=1$.

Fig.\ref{fig.1} presents the QFI values for a single-qubit pseudo-Hermitian system and its corresponding Hermitian system, derived using the Naimark dilation extension method.  The QFI $F_x(n)$ is influenced by the parameter $n$ which arises from the biorthonormal property of the non-Hermitian eigenstates. In the equivalent Hermitian system obtained through the dilation method, the precision of the estimated parameter is improved, which is consistent with the findings reported in Ref. \cite{PhysRevLett.131.160801}. It is noteworthy that the Hermitian system, derived through Naimark dilation of the original non-Hermitian system, generally retains the QFI of the original system. This dilation process is not merely an extension of the Hilbert space but rather involves incorporating an auxiliary system to extend the evolution information of the non-Hermitian system into a larger Hermitian framework. During this embedding procedure, as much information as possible is preserved from the original system. Consequently, within the enlarged Hilbert space, measurements using Hermitian operators tend to yield more accurate parameter estimates. As discussed previously, the QFI $F^d_x$ for the eigenstate $|\psi_2\rangle$ in the extended Hermitian system is higher than the QFI $F_x\left(n=1/\sqrt{1+\delta_{\lambda}^{2}}\right)$ for the projective state $|\psi_1\rangle$. However, increasing the value of $n$ introduces more informational content to the QFI. These observations validate the feasibility and correctness of the non-Hermitian QFI definition we have proposed. In this regard, the non-Hermitian QFI seems to be appropriately defined by the projective state $|\psi_1\rangle$ independently of the non-Hermiticity parameter  $\alpha$ , as also discussed in Ref. \cite{PhysRevA.108.022215}.

However, it is important to note that in non-Hermitian systems, the evolution operators typically possess complex eigenvalues, which leads to non-normalizable states. As a result, the non-Hermiticity parameter $\alpha$, which is tied to the normalization factor, contains information about the system parameters. This factor, as elaborated in our definition, provides additional insights into the parameters and contributes to the enhancement of the QFI. Moreover, during the time evolution of the system, exponential gain or dissipation can occur, which makes the system state highly sensitive to small variations in the parameters, further increasing the QFI. Next, we will examine the influence of non-Hermiticity on dynamic evolution and QFI.

\section{$\mathcal{PT}$-SYMMETRIC HAMILTONIAN SYSTEM}\label{sec4}
Now, we consider a two-level system governed by a $\mathcal{PT}$-symmetric Hamiltonian \cite{2017Experimental,article2019} which can be expressed as
\begin{equation}\label{Eq.36}
		H_{s}=
		\left(
		\begin{array}{cc}
			re^{i\omega} & s \\
			s & re^{-i\omega}
		\end{array}
		\right),
	\end{equation}
 where $r$, $s$, and $\theta$ are real. The eigenvalues are given by
 \begin{equation}
 \varepsilon_{\pm}=r\cos\omega\pm\sqrt{s^{2}-r^{2}\sin^{2}\omega}
\end{equation}
When $s^{2}> r^{2}{\sin}^{2}\omega$, the eigenvalues $\varepsilon_\pm$ are real, the system is in the unbroken $\mathcal{PT}$-symmetric region. Conversely, when $s^{2}< r^{2}\sin^{2}\omega$, the eigenvalues become complex, indicating  the broken $\mathcal{PT}$-symmetric region. At the exceptional point (EP), where $s^{2}=r^{2}\sin^{2}\omega$, the eigenvalues coalesce: $\varepsilon=r\cos\omega$.

 We first consider the the unbroken $\mathcal{PT}$-symmetry phase. In this region, the energy eigenstates corresponding to $\varepsilon_\pm$ are
 \begin{subequations}
      \begin{equation}\label{Eq.37a}
		|\varepsilon_{+}\rangle=\frac{1}{\sqrt{2}}
		\left(
		\begin{array}{c}
			e^{ix/2}  \\
			e^{-ix/2}
		\end{array}
		\right),
	\end{equation}
 and
 \begin{equation}\label{Eq.37b}
		|\varepsilon_{-}\rangle=\frac{i}{\sqrt{2}}
		\left(
		\begin{array}{c}
			e^{-ix/2}  \\
			-e^{ix/2}
		\end{array}
		\right),
	\end{equation}
 \end{subequations}
where $\sin x\equiv(r/s)\sin\omega$. Due to the non-Hermicity, these eigenstates are not orthogonal, i.e., $\langle\varepsilon_{+}|\varepsilon_{-}\rangle=\sin x$.  By the Hamiltonian in Eq. (\ref{Eq.36}),  the evolution operator can be expressed as \cite{PhysRevLett.89.270401}
 \begin{equation}\label{Eq.38}
  \begin{aligned}
     U_{\theta}&=e^{-iH_{s}\theta}\\
     &=\frac{1}{\cos x}
     \left(
     \begin{array}{cc}
     \cos(\theta^{'}-x) & -i\sin(\theta^{'}) \\
     -i\sin(\theta^{'}) & \cos(\theta^{'}+x)
     \end{array}
     \right),
     \end{aligned}
 \end{equation}
where $\theta^{'}=s\theta \cos x$. For any nonzero evolution parameter $\theta$, the off-diagonal elements evolution operator of $U_\theta$ are purely imaginary, indicating that the evolution is non-unitary. For convenience, we choose $\theta=\pi/(2s\cos x)$, which simplifies the evolutions operator to
 \begin{equation}\label{Eq.39}
     U_{\theta}=\frac{1}{\cos x}
     \left(
     \begin{array}{cc}
        \sin x  & -i \\
        -i  & -\sin x
     \end{array}
     \right).
\end{equation}
It is easy to see that this non-unitarity of the evolution depends solely on the parameter $x$ . As $x$ approaches the EP point, the two eigenstates coalesce. When $x$ equals to $0$, the $\mathcal{PT}$-symmetric case reduces to a conventional quantum mechanics case.

We consider an arbitrary initial state $|\psi_{0}\rangle=N(|\varepsilon_{+}\rangle+me^{i\phi}|\varepsilon_{-}\rangle)$. The normalization factor is given by $1/N^{2}=1+m^{2}+2m\sin x\cos\phi$.  It is worth to noting that the eigenstates are normalized; therefore, the non-Hermiticity arises solely from the inner product factor, which can be derived as
\begin{equation}\label{Eq.40}
\begin{aligned}
P_{\theta}&=\langle\psi_{0}|\hat{U}_{\theta}^{\dagger}\hat{U}_{\theta}|\psi_{0}\rangle\\
&=N^{2}(e^{i(\varepsilon_{+}^{*}-\varepsilon_{+})\theta}+m^{2}e^{i(\varepsilon_{-}^{*}-\varepsilon_{-})\theta}\\
&+m\sin x e^{i\phi}e^{i(\varepsilon_{+}^{*}-\varepsilon_{-})\theta}+m\sin x e^{-i\phi}e^{i(\varepsilon_{-}^{*}-\varepsilon_{+})\theta})\\
&=N^{2}(1+m^{2}+2m\sin x\cos\varphi)\\
&=e^{2\alpha_\theta},
\end{aligned}
\end{equation}
where $\varphi=(2s\cos x)\theta+\phi$. This shows that the non-Hermiticity factor $\alpha_\theta$ depends with $\theta$, and vanishes when $\theta=0$,  indicating no non-Hermitian effect on the QFI at the initial state $|\psi_0\rangle$ . However, even if  $|\psi_\theta\rangle$ is initially prepared as a normalized projective state. The evolution governed by $U_\theta$ introduces non-Hermiticity as  $\theta$ increases. Consequently, the influence of the non-Hermiticity factor $\alpha_\theta$ on both the state evolution and the QFI becomes significant.

We further compute the expectation values:
\begin{subequations}
    \begin{equation}\label{Eq.41a}
    \begin{aligned}
        \langle H_{s}^{\dagger}H_{s}\rangle_{\theta}&=N^{2}\{\varepsilon_{+}^{2}e^{i(\varepsilon_{+}^{*}-\varepsilon_{+})\theta}+\varepsilon_{-}^{2}m^{2}e^{i(\varepsilon_{-}^{*}-\varepsilon_{-})\theta}\\
        &+2Re[m\varepsilon_{+}\varepsilon_{-}\sin x e^{i\phi}e^{i(\varepsilon_{+}^{*}-\varepsilon_{-})\theta}]\}\\
        &=N^{2}(\varepsilon_{+}^{2}+m^{2}\varepsilon_{-}^{2}+2m\varepsilon_{+}\varepsilon_{-}\sin x\cos\varphi)
    \end{aligned}
\end{equation}
\begin{equation}\label{Eq.41b}
    \begin{aligned}
        \langle H_{s}\rangle_{\theta}&=N^{2}(\varepsilon_{+}e^{i(\varepsilon_{+}^{*}-\varepsilon_{+})\theta}+\varepsilon_{-}m^{2}e^{i(\varepsilon_{-}^{*}-\varepsilon_{-})\theta}\\
        &+m\sin x(\varepsilon_{+}e^{i\phi}e^{i(\varepsilon_{+}^{*}-\varepsilon_{-})\theta}+\varepsilon_{-}e^{-i\phi}e^{i(\varepsilon_{-}^{*}-\varepsilon_{+})\theta})\\
        &=N^{2}[(r\cos\omega+s\sin x)+m^{2}(r\cos\omega-s\sin x)\\
        &+2m\sin x(r\cos\omega\cos\varphi+is\cos x\sin\varphi)]
    \end{aligned}
\end{equation}
\end{subequations}

According to Eq.(\ref{Eq.17}), the QFI under the evolution governed by $U_\theta$ becomes
\begin{equation}\label{Eq.42}
\begin{aligned}
    F_{\theta}^{r}&=\frac{4(\partial_{\theta}P_{\theta})^{2}}{P_{\theta}}+4 (\langle H_{s}^{\dagger}H_{s}\rangle_{\theta}-\langle H_{s}^{\dagger}\rangle_{\theta}\langle H_{s}\rangle_{\theta} )\\
    &=\frac{64N^{2} (m^2 \sin ^2\alpha  \cos ^2\alpha \sin ^2\varphi)}{(m^2+2 m \sin\alpha\cos\phi+1)}\\
     &+\frac{16N^2m^{2}(s^{2}-r^{2}\sin^{2}\omega)^{2}}{(s+sm^{2}+2mr\sin\omega\cos\varphi)^{2}}
\end{aligned}
\end{equation}
The value of  QFI is sensitive to the choice of initial state. Therefore, it is essential to determine the optimal initial state to maximize the QFI. The initial state $|\psi_0\rangle$, which is set to be normalized ($\langle\psi_0|\psi_0\rangle=1$), depends on the parameters $m$ and $\phi$, and the optimal values of these parameters can be determined by solving $\partial F_{\theta}^{r}/\partial m=0$ and $\partial F_{\theta}^{r}/\partial\phi=0$. Setting $\partial F_{\theta}^{r}/\partial m=0$ yields extremum points are obtained at $m=\pm 1$ and$ m=0$. While $\partial F_{\theta}^{r}/\partial\phi=0$ gives $\phi=(2k+1)\pi$, where $ k\in Z$. The QFI reaches its maximum when $m = \pm 1$ and $\phi=\pi$. Consequently, the optimal initial state in the unbroken-$\mathcal{PT}$-symmetric region is given by
\begin{equation}
    |\psi_{0}\rangle=N(|\varepsilon_{+}\rangle\pm e^{i\pi}|\varepsilon_{-}\rangle)
\end{equation}.
\begin{figure}[ht]
	\centering
	\includegraphics[width=0.41\textwidth]{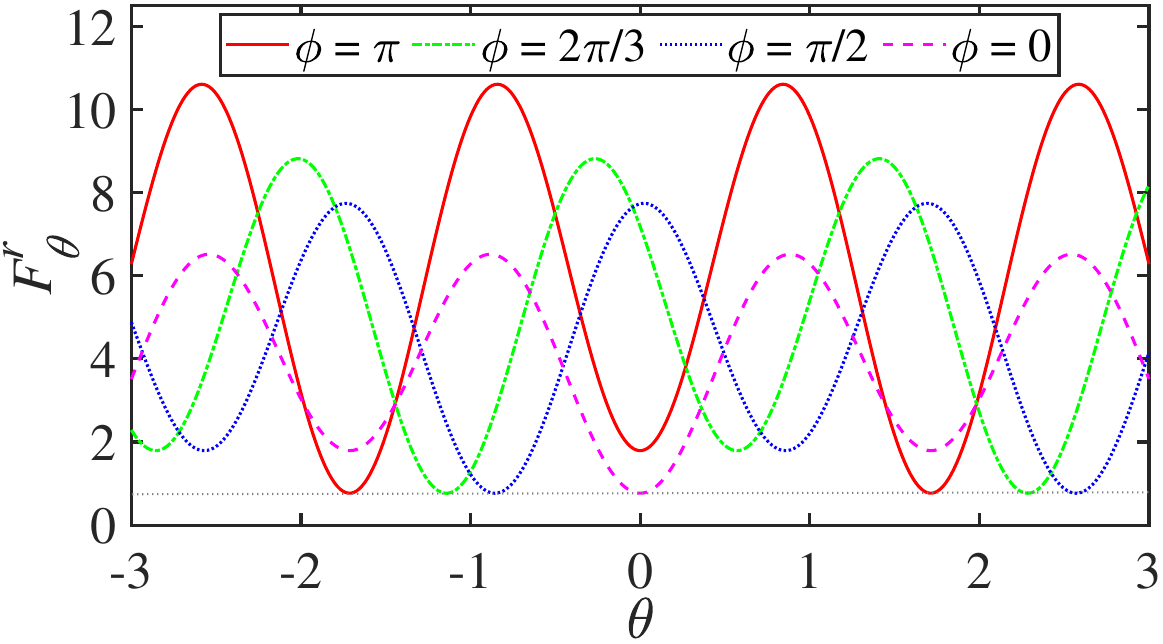}
	\caption{Evolution of the QFI $F_{\theta}^{r}$ as a function of the parameter $\theta$ in the unbroken  $\mathcal{PT}$-symmetric region. The other parameters are chosen as \( m = 1 \), \( s = 1 \), \( r = 0.4 \), \( \omega = \pi/2 \), and \( \phi = \pi, 2\pi/3, \pi/2, 0 \). The red solid line, green dash-dot line, blue dotted line and purple dashed line correspond to \( \phi = \pi, 2\pi/3,\pi/3\), and \( \phi = 0 \), respectively.}
 \label{fig.2}
\end{figure}
As shown in Fig.\ref{fig.2}, in the unbroken $\mathcal{PT}$-symmetric region, the QFI exhibits an oscillatory behavior with respect to \(\theta\). In the measurement process, we aim to maximize the QFI corresponding to a specific parameter \(\theta_0\). Since the QFI varies with \(\theta\), the optimal initial state for maximizing the QFI also depends on \(\theta\). By varying the value of \(\phi\), we can shift the curve along the \(\theta\) axis, and the value of \(\phi\) also influences the magnitude of the QFI. We can maximize the QFI for the specific parameter \(\theta_0\) by adjusting \(\phi\).
\begin{figure}[ht]
	\centering
	\includegraphics[width=0.41\textwidth]{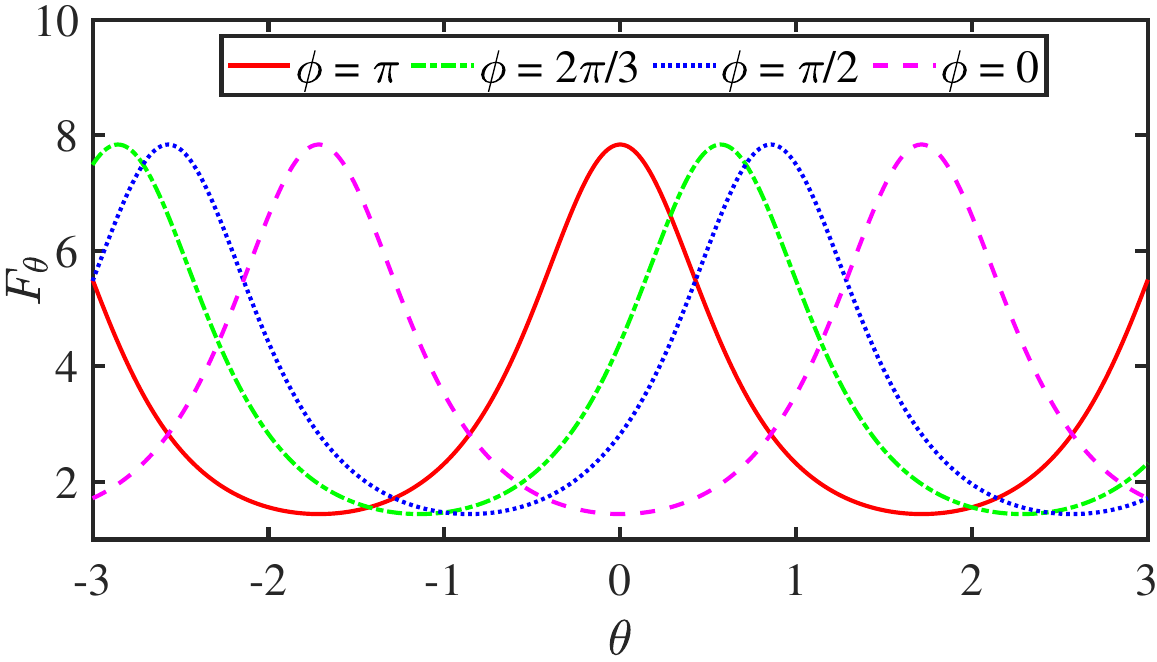}
	\caption{Evolution of the QFI $F_{\theta}^{r}$ as a function of the parameter $\theta$ calculated by using the projected quantum states. The other parameters are chosen as the same as Fig. \ref{fig.2}}
 \label{fig.3}
\end{figure}

When calculating the QFI using projected quantum states without accouting for normalization coefficient, the variation of QFI with the parameter is shown in Fig.\ref{fig.3}. It is evident that due to the non-Hermicity, the evolution operator $U_{\theta}$ is non-unitary. The QFI is influenced by the non-Hermicity parameter $\alpha$. Comparing Fig.\ref{fig.2} with Fig.\ref{fig.3} reveals that $\alpha$ affects the QFI maximum and oscillation periods. Moreover, the extent of this influence is dependent on the choice of the initial state. When $\phi=\pi,0$, the impact is most pronounced. Specifically, for the initial state with $\phi=\pi$, the non-Hermicity enhances the QFI. This indicates that the non-Hermicity parameter $\alpha$ encodes valuable information about the parameter to be estimated, and incorporating $\alpha$ into the QFI analysis can improve the precision of parameter estimation.

In the broken $\mathcal{PT}$-symmetric region, the eigenvalues of the Hamiltonian become complex and are given by $\lambda_{\pm}=r\cos\omega\pm i\sqrt{r^{2}\sin^{2}\omega-s^{2}}$, while the corresponding eigenstates take the form
\begin{equation}\label{Eq.43}
    |\lambda_{\pm}\rangle=\frac{1}{\sqrt{2r\sin\omega}}
    \left(
    \begin{array}{c}
          i\sqrt{r\sin\omega\pm\kappa} \\
         \sqrt{r\sin\omega\mp\kappa}
    \end{array}
    \right),
\end{equation}
where $\kappa=\sqrt{r^{2}\sin^{2}\omega-s^{2}}$. Similarly, these eigenstates are set to be normalized $\langle\lambda_{\pm}|\lambda_{\pm}\rangle=1$, with the overlap given by $\langle\lambda_{+}|\lambda_{-}\rangle =\mu=s/r\sin\omega$. For an arbitrary normalized initial state $|\psi_{0}\rangle=N(|\lambda_{+}\rangle+me^{i\phi}|\lambda_{-}\rangle)$, the normalization condition $\langle\psi_{0}|\psi_{0}\rangle=1$ leads to $1/N^{2}=1+m^{2}+2m\mu\cos\phi$. Consequently, we can derive the QFI as

\begin{equation}\label{Eq.44}
\begin{aligned}
    F_{\theta}^{i}&=\frac{16N^{2}\kappa^{2}(e^{2\kappa\theta}-m^{2}e^{-2\kappa\theta})^{2}}{(e^{2\kappa\theta}+m^{2}e^{-2\kappa\theta}+2m\mu\cos\phi)}\\
    &+\frac{16m^{2}\mu^{2}(r^{2}\sin^{2}\omega-s^{2})^{2}e^{4\kappa\theta}}{[s(e^{4\kappa\theta}+m^{2})+2ms\mu e^{2\kappa\theta}\cos\phi]^{2}}.
    \end{aligned}
\end{equation}

In the broken $\mathcal{PT}$-symmetry region, this expression for the QFI indicates that both the choices of $m$ and $\phi$ influence the QFI. To determine the optimal initial state, we solve the conditions $\partial F_{\theta}^{i}/\partial m = 0$ and $\partial F_{\theta}^{i}/\partial \phi = 0$. This analysis leads to the conclusion that the optimal initial state is
\begin{equation}\label{Eq.45}
    |\psi_{0}\rangle=N(|\lambda_{+}\pm e^{i\pi}|\lambda_{-}\rangle),
\end{equation}
corresponding to $m=\pm 1$ and $\phi=\pi$. In this case, the QFI no longer contains oscillatory components; instead, it features exponential amplification or attenuation effects.   As shown in Fig.\ref{fig.3}, in the corresponding unbroken $\mathcal{PT}$ symmetric region, the QFI oscillates with respect to \(\theta\). In contrast, when $\mathcal{PT}$ symmetry is broken, the QFI exhibits rapid decay or gain as \(\theta\) changes. Variations in the overlap coefficient $\mu$ lead to different peak or valley structures in the QFI.  These findings highlight that the QFI performs significantly better in the unbroken $\mathcal{PT}$-symmetric regime, making it more favorable for high-precision parameter estimation.

\begin{figure}[ht]

        \centering
        \includegraphics[width=\linewidth]{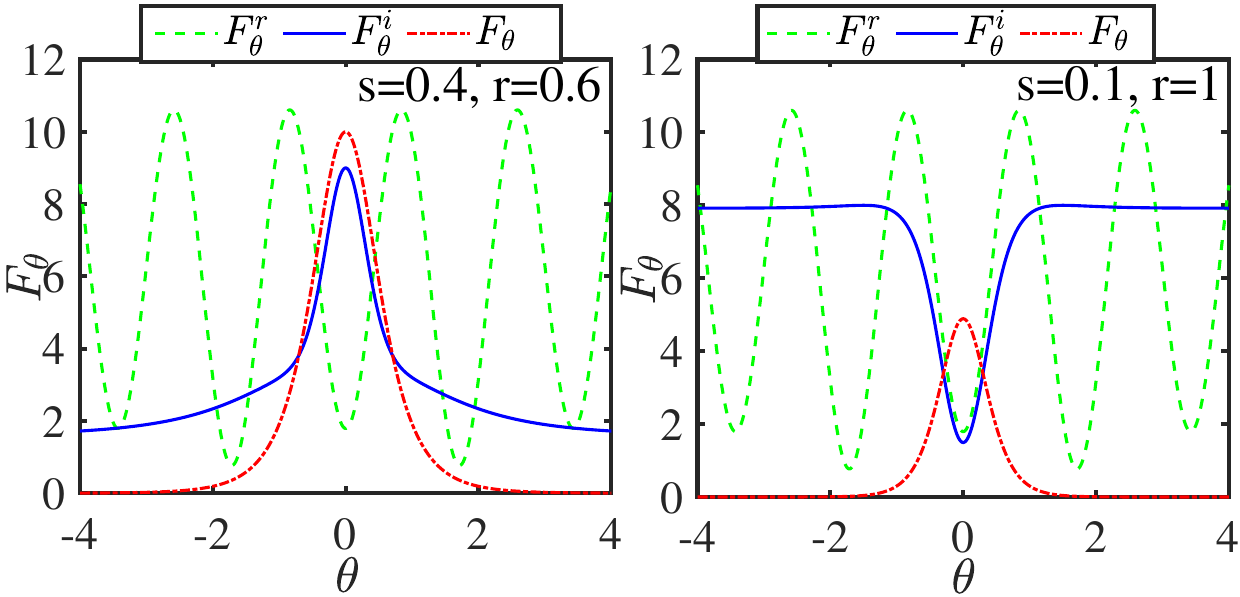}

    \caption{The green dashed line represents the variation of \(F_{\theta}^{r}\) with respect to \(\theta\) in the unbroken $\mathcal{PT}$-symmetric region. The blue solid line illustrates the variation of \(F_{\theta}^{i}\) with respect to \(\theta\) in the broken $\mathcal{PT}$-symmetry region. The red dash-dot line represents the variation of $F_{\theta}$ with respect to the parameter $\theta$ calculated by using the projected quantum states. In the $\mathcal{PT}$-symmetric region, the initial state is $N(|\varepsilon_{+}\rangle+e^{i\pi}|\varepsilon_{-}\rangle$ and the parameters are set to \(s = 1\), \(r = 0.4\), and \(\omega = \pi / 2\). In the $\mathcal{PT}$-broken region, the initial state is $N(|\lambda_{+}\rangle+e^{i\pi}|\lambda_{-}\rangle$. (Left panel) We set \( s = 0.4 \), \( r = 0.6 \), and \( \omega = \pi / 2 \). The value of the QFI is approximately 8, showing a rapid decline in the region where \(\theta\) approaches 0. (Right panel) We set \( s = 0.1 \), \( r = 1 \), and \( \omega = \pi / 2 \). Conversely, as \(\theta\) approaches 0, there is a rapid increase that leads to a peak. }

    \label{fig.4}
\end{figure}

As for EP, the eigenvalues are degenerate $\lambda=r\cos\omega$. If the initial state is on the eigenstates, we have
\begin{equation}\label{Eq.46}
\begin{aligned}
    F_{\theta}&=4(\langle\psi_{0}|e^{i\lambda^{*}\theta}\lambda^{*}\lambda e^{-i\lambda\theta}|\psi_{0}\rangle-|\langle\psi_{0}|e^{i\lambda^{*}\theta}\lambda e^{-i\lambda\theta}|\psi_{0}\rangle|^{2})\\
    &=4(\lambda^{2}-\lambda^{2})=0
\end{aligned}
\end{equation}
Although the value of the QFI at the EP is zero, the computation involves the eigenstates at the EP. At the EP, the eigenstates are degenerate, and thus do not form a complete basis set in the state space. Therefore, it is possible that the QFI at the EP is not actually zero.

\section{CONCLUSIONS}\label{sec5}
To derive the expression of Quantum Fisher Information (QFI) within a non-Hermitian framework, this work begins with the Schwarz inequality and introduces a novel definition of the non-Hermitian symmetric logarithmic derivative. By employing the projected Hilbert space method tailored for non-Hermitian systems, we obtain the QFI for single-parameter estimation and establish the corresponding quantum Cramér-Rao bound. Furthermore, using spectral decomposition and parameter-dependent evolution, we provide an explicit expression for the QFI in terms of the density matrix and parameter generators.  This definition of non-Hermitian QFI for states in non-Hermitian systems not only aligns with the results in the Hermitian case, but also reveals the non-Hermitian effects arising from the time-dependent norm of the state.

To illustrate our results, we analyze parameter estimation in a single-qubit pseudo-Hermitian system. Utilizing Naimark dilation theory and introducing an auxiliary system, the pseudo-Hermitian system is projected into an equivalent enlarged Hermitian system. A comparative analysis of the QFI before and after dilation confirms the validity of our formulation and demonstrates the potential to enhance parameter estimation precision. In contrast to a standard Hermitian system without additional parameters, the system obtained via Naimark dilation retains more information about the estimated parameter.

To discuss the influence of the non-Hermiticity brought by the time-dependent Norm of the state, we have also studied a PT-symmetric system. The parameter estimation performance has been superior in the PT-symmetric region compared to the PT-broken region. Optimal initial states for maximizing QFI have been determined, and adjusting the relative phase $\phi$ of the initial state can yield the most suitable state for estimating specific parameters. Furthermore, in the PT-broken region, QFI has been highly sensitive to parameter changes, making it suitable for application in the quantum sensing field. Considering the importance of Quantum Fisher Information and PT-symmetric systems in optics and quantum information, the QFI of non-Hermitian Hamiltonians shows promising potential for advancing quantum metrology and the development of non-Hermitian sensing.
\begin{acknowledgments}
This work is supported by the National Natural Science Foundation of China (NSFC) (Grants No. 12147206, No. 12088101, and No. U2330401) and the Science and Technology Development Plan Project of Jilin Province, China (Grants No. 20240101321JC).
\end{acknowledgments}

\end{document}